%% file: ms.tex
\documentclass[journal]{IEEEtran}
\usepackage{cite}
\usepackage{amsmath}
\usepackage{amsthm}
\usepackage{algorithmic}
\usepackage{url}

\usepackage{amsfonts}

\usepackage{subcaption}
\usepackage{psfrag}
\usepackage{pstricks}
\usepackage{pst-plot}
\usepackage{pst-circ}
\usepackage{enumitem}
\usepackage{graphicx}
\usepackage[short]{optidef}

\graphicspath{{./pics/}}
\DeclareGraphicsExtensions{.eps}

\input{mycommands}

\newcommand{\gen}[1]{g^{(#1)}}

\newcommand{\fset}[0]{\mathcal{P}}

\newsavebox{\ieeealgbox}
\newenvironment{boxedalgorithmic}
  {\begin{lrbox}{\ieeealgbox}
   \begin{minipage}{\dimexpr\columnwidth-2\fboxsep-2\fboxrule}
   \begin{algorithmic}}
  {\end{algorithmic}
   \end{minipage}
   \end{lrbox}\noindent\fbox{\usebox{\ieeealgbox}}}

\begin{document}

%
\title{Aggregation and Disaggregation of Energetic Flexibility from Distributed Energy Resources}
%
%
%

\author{Fabian~L.~M\"uller,
		J\'acint~Szab\'o,
		Olle~Sundstr\"om,
        and~John~Lygeros,~\IEEEmembership{Fellow,~IEEE}
%
\thanks{F. L. M\"uller is with the Automatic Control Laboratory, Swiss Federal Institute of Technology, Zurich, Switzerland, and IBM Research--Zurich, R\"uschlikon, Switzerland.
        {\tt\small fmu@zurich.ibm.com}}%
\thanks{J. Szab\'o and O. Sundstr\"om are with IBM Research--Zurich, R\"uschlikon, Switzerland.
        {\tt\small \{jsz,osu\}@zurich.ibm.com}}%
\thanks{J. Lygeros is with the Automatic Control Laboratory, Swiss Federal Institute of Technology, Zurich, Switzerland.
        {\tt\small jlygeros@ethz.ch}}%
}

\maketitle

\input{abstract}


\input{introduction}
\input{flexDescription}
\input{zonoApprox}
\input{aggregation}

\input{disaggregation}

\input{zonoUseCases}
\IEEEpeerreviewmaketitle

\input{conclusion}


%

\input{appendix}

\section*{Acknowledgment}
The authors gratefully acknowledge the fruitful discussions with Stefan W\"orner, Ulrich Schimpel, and Marco Laumanns.


\ifCLASSOPTIONcaptionsoff
  \newpage
\fi


\bibliographystyle{IEEEtran}
\bibliography{IEEEabrv,../../../Papers/BibTeX/library}

\vfill
\begin{IEEEbiographynophoto}
{Fabian M\"uller} received the B.Sc. degree in mechanical engineering and the M.Sc. degree in robotics, systems and control from the Swiss Federal Institute of Technology (ETH) Zurich, Switzerland in 2009 and 2012, respectively. 

He was recently affiliated with the Institute for Dynamic Systems and Control at ETH Zurich. 
 
Mr. M\"uller is currently a doctoral student with the Automatic Control Laboratory at ETH Zurich and with IBM Research Zurich. His research interests include technologies for smart power grids, control theory and optimization.
\end{IEEEbiographynophoto}
\vspace*{-2\baselineskip}
\begin{IEEEbiographynophoto}
{J\'acint Szab\'o} received the M.Sc. degree in mathematics in 2002, and the Ph.D. degree in mathematics in 2007, both from E\"otv\"os University, Budapest.

He worked as an Associate Professor in the Informatics Laboratory of the Computer Science Institute, Budapest, and in the Mathematical Optimization Department of the Zuse Institut, Berlin. 

Since 2011, Dr. Szab\'o is a Research Staff Member in the Cognitive Computing and Industry Solutions department at IBM Research Laboratory, Zurich, Switzerland. His research interests include optimization in transportation and smart grids.
\end{IEEEbiographynophoto}
\vspace*{-2\baselineskip}
\begin{IEEEbiographynophoto}
{Olle Sundstr\"om} was born in \"Orebro, Sweden, in 1980. He received the M.Sc. degree in electrical engineering from Chalmers University of Technology, Gothenburg, Sweden, in 2006, and the Doctor of Sciences degree from the Swiss Federal Institute of Technology, Zurich, Switzerland, in 2009. 

He was previously a Research Assistant at the Institute for Dynamic Systems and Control at the Swiss Federal Institute of Technology, Zurich, Switzerland, as well as the Powertrain Research Group, Empa, Swiss Federal Laboratories for Materials Testing and Research, D\"ubendorf, Switzerland. 

Dr. Sundstr\"om is currently a Research Staff Member in the Cognitive Computing and Industry Solutions department at IBM Research Laboratory, Zurich, Switzerland. His research interests include smart grids, demand side management, optimal control and optimization, especially for applications of energy management of plug-in electric vehicle fleets.
\end{IEEEbiographynophoto}
\vspace*{-2\baselineskip}
\begin{IEEEbiographynophoto}
{John Lygeros} (S'91-–M'97-–SM'06-–F'11) received the B.Eng. degree in electrical engineering and the M.Sc. degree in systems control from Imperial College of Science Technology and Medicine, London, U.K., in 1990 and 1991, respectively, and the Ph.D. degree from the Electrical Engineering and Computer Sciences Department, University of California, Berkeley, CA, USA, in 1996.

He holds the chair of Computation and Control at the Swiss Federal Institute of Technology (ETH) Zurich, Switzerland, where he is currently serving as the Head of the Automatic Control Laboratory. During the period 1996--2000, he held a series of research appointments at the National Automated Highway Systems Consortium, Berkeley, the Laboratory for Computer Science, M.I.T., and the Electrical Engineering and Computer Sciences Department at U.C. Berkeley. Between 2000 and 2003, he was a University Lecturer at the Department of Engineering, University of Cambridge, U.K., and a Fellow of Churchill College. Between 2003 and 2006, he was an Assistant Professor at the Department of Electrical and Computer Engineering, University of Patras, Greece. In July 2006, he joined the Automatic Control Laboratory at ETH Zurich, first as an Associate Professor, and since January 2010 as a Full Professor. His research interests include modeling, analysis, and control of hierarchical, hybrid, and stochastic systems, with applications to biochemical networks, transportation, power grids and camera networks.

Dr. Lygeros is a member of the IET and of the Technical Chamber of Greece.
\end{IEEEbiographynophoto}







\end{document}

%% file: mycommands.tex
\newcommand{\fig}{Fig.\,}
\newcommand{\tab}{Tab.\,}
\newcommand{\sect}{Sec.\,}

\newcommand{\cf}{cf.\,}
\newcommand{\ie}{i.e.\,}
\newcommand{\eg}{e.g.\,}

\newcommand{\un}[1]{\underline{#1}}

\newcommand{\R}[1]{\mathbb{R}^{#1}}

\newcommand{\e}{\begin{equation}}
\newcommand{\ee}{\end{equation}}
\newcommand{\trans}{^{\top}} 
\newcommand{\opt}{^{\ast}} 
\newcommand{\ind}[1]{^{(#1)}}

\newcommand{\argmax}{\operatornamewithlimits{argmax}}

\newcommand{\ts}[1]{\text{#1}}

\newtheorem{thm}{Theorem}
\newtheorem{clm}{Claim}


%% file: abstract.tex
\begin{abstract}
A variety of energy resources has been identified as being flexible in their electric energy consumption or generation. This energetic flexibility can be used for various purposes such as minimizing energy procurement costs or providing ancillary services to power grids. To fully leverage the flexibility available from distributed small-scale resources, their flexibility must be quantified and aggregated.
This paper introduces a generic and scalable approach for flexible energy systems to quantitatively describe and price their flexibility based on zonotopic sets. The description proposed allows aggregators to efficiently pool the flexibility of large numbers of systems and to make control and market decisions on the aggregate level. In addition, an algorithm is presented that distributes aggregate-level control decisions among the individual systems of the pool in an economically fair and computationally efficient way. Finally, it is shown how the zonotopic description of flexibility enables an efficient computation of aggregate regulation power bid-curves.

\end{abstract}

%% file: introduction.tex
\section{Introduction}
\label{s:introduction}
\IEEEPARstart{I}{ncreasing} shares of distributed renewable energy resources lead to a growing need for energetic flexibility in power grids. 
On the one hand, renewable energy sources, such as wind and solar power, introduce large amounts of variability and uncertainty to power grids because of their inherent intermittency and limited controllability. 
On the other hand, substituting well-controllable traditional generating units, such as nuclear, gas, and pumped-hydro storage power plants, by renewables reduces the rotational inertia of the power system and the supply of highly reliable regulation power \cite{Ulbig2014b}. Consequently, future power grids will require additional sources of flexibility to guarantee safe and reliable operation. 

The flexibility of distributed, small-scale energy systems, on both the demand and the supply side, can serve as an additional resource to the large-scale supply flexibility that has been traditionally used to balance the grid. Sources of small-scale flexibility include heating, ventilation, and air-conditioning systems, plug-in electric vehicles and stationary batteries, as well as micro-generation units. However, quantifying and making use of the flexibility of large numbers of distributed systems is challenging, in particular because of the computational complexity. To take full advantage of the flexibility available and to reduce the complexity of planning, trading and control, the aggregate flexibility of an entire group of systems must be computed and represented in a concise and compact form. The importance of aggregation as a key enabler for incorporating large numbers of flexible systems has been highlighted in numerous works, \eg, \cite{Eid2015,Gkatzikis2013,Trangbaek2012,Nayyar2013,Zhao2016b}. The tasks of collecting, aggregating and controlling the flexibility of a group of systems are performed by an entity called \emph{aggregator} \cite{Koponen2012,Barot2014}. The aggregator concludes contracts with individual flexible energy resources, defining the way of communication and control between the systems and the aggregator, the specifics of the flexibility offered by the systems, as well as the details on how the systems are reimbursed for the flexibility provided \cite{Koponen2012,Eid2015,Gkatzikis2013,Balandat2014}. Thus, the aggregator acts as an intermediary between flexible resources and wholesale markets, molding individual flexibilities into tradable products. 


Various approaches have been proposed to characterize the energetic flexibility both of individual systems and of entire populations. 
We consider flexibility from the perspective of the power grid and define it as the \emph{ability of a system to adjust the volume and/or timing of its electric energy intake from or of its energy output to the power grid}. Consequently, the set of all power trajectories that can be tracked by the system over a given time horizon is a natural description of its flexibility and is referred to as the \emph{feasible set} \cite{Trangbaek2012,Mueller2015}. 
Among the approaches for characterizing the feasible set of individual systems are \emph{power nodes} \cite{Ulbig2012,Koch2012}, \emph{generalized batteries} \cite{Hao2013,Nayyar2013,Alizadeh2014}, and \emph{resource polytopes} \cite{Trangbaek2012,Mueller2015,Zhao2016b}. 

Regarding the description of the flexibility of an entire population of systems, two main types of approaches can be identified. Top-down approaches attempt to capture the aggregate flexibility directly based, for example, on the probabilistic properties of the underlying systems \cite{Koch2011,Kamgarpour2013,Sajjad2016}. In contrast, here we consider a bottom-up approach in which the flexibility of each individual system is described first and aggregated subsequently. The particular description of individual feasible sets determines the computational complexity of aggregating them. In general, computing the exact aggregate feasible set is computationally intractable \cite{Trangbaek2012,Zhao2016b}. Consequently, attempts to compute the exact aggregate feasible set are restricted to special cases, such as symmetric constraints \cite{Hao2013} or binary resource polytopes \cite{Trangbaek2012}.
Other approaches use set descriptions that provide inner- and/or outer-approximations of the exact aggregate feasible set. Outer-approximations are derived based on polytopes \cite{Barot2014,Gonzalez2015} or semi-definite constraints \cite{Barot2016}. While some outer-approximations can be computed efficiently, the fact that they can contain infeasible power trajectories is a major drawback. In contrast, inner-approximations provide sufficient conditions for feasibility. Approaches for computing inner-approximations are based on zonotopics sets \cite{Mueller2015} or on projection to polytopic sets \cite{Zhao2016b}. Both inner- and outer-approximations for battery models are provided in \cite{Alizadeh2014,Hao2013}. 

Compared to flexibility aggregation, disaggregation, \ie, distributing an aggregate control signal among the systems of a population, has attracted limited attention. Hierarchical control schemes have been proposed in \cite{Trangbaek2011,Trangbaek2012}. Other approaches use a priority-stack \cite{Hao2013} or randomized \cite{Kamgarpour2013} controllers. 

The main contribution of this work is to introduce zonotopes, a subclass of polytopes, as approximations of the original feasible set of a flexible system. Zonotopes can be aggregated efficiently with regard to memory requirements and computational complexity and can capture time-variant, asymmetric power, energy and ramp constraints. 
In addition, an efficient disaggregation algorithm based on zonotopes is presented that distributes a given aggregate-level signal among the systems in an economically fair way. Moreover, an aggregator can use the algorithm to compute regulation power bids.

Section \ref{s:flexDescription} introduces the idea of using zonotopes to approximate the flexibility of individual systems. Optimal zonotopic approximations are computed in \sect\ref{s:zonoApprox}. \sect\ref{s:aggregation} discusses the aggregation of flexibility, followed by \sect\ref{s:disaggregation} where a disaggregation algorithm based on flexibility costs is derived. In \sect\ref{s:zonoUseCases}, zonotopic descriptions of flexibility are used to perform feasibility checks and to compute bid curves for regulation power. A conclusion and outlook are provided in \sect\ref{s:conclusion}.

%% file: flexDescription.tex
\section{Description of flexibility}
\label{s:flexDescription}



\subsection{Polytopic feasible sets}
\label{ss:polytopicFeasSets}
Consider a single energy resource over a finite discrete-time horizon comprising $N$ time steps each of duration $t_s$. By definition, the feasible set ${\fset\subseteq\R{N}}$ comprises all the power trajectories ${p\in\R{N}}$ the system is able to follow. This set is defined by the dynamics and constraints the system must respect. The constraints most common to flexible energy resources are the following.

	\textit{1) Power constraints:} In general, the constant power $p_k$ the system draws from (${p_k>0}$) or feeds into (${p_k<0}$) the power grid during time step $k$, is bounded, \ie,
\begin{align}
	\un{p}_k &\leq p_k\leq \bar{p}_k,\ k=1,\dots,N.
	\label{eq:powerConstr}
\end{align}

	\textit{2) Energy constraints:} Most flexible energy resources owe their flexibility to an energy buffer with limited capacity. Thus, the cumulative electric energy $e_k$ the system can consume or generate up to and including time slot $k$ is bounded, \ie,
\begin{equation}
	\un{e}_k-e_0 \leq t_s \sum\limits_{i=1}^k p_i\leq \bar{e}_k-e_0,\ k=1,\dots,N,
\label{eq:energyConstr}
\end{equation}
where $e_0$ is the initial energy level of the system. 

	\textit{3) Ramp-rate constraints:} Often, there are limits on the rate at which power can vary, \ie,
\begin{equation}
	\un{r}_k \leq (p_k-p_{k-1})/t_s \leq \bar{r}_k,\ k=2,\dots,N.
	\label{eq:rampConstr}
\end{equation}

	\textit{4) State constraints:} Various energy resources have internal states that are driven by the electric power consumption or generation of the system. Examples are the state-of-charge of a battery or the temperature of a heating system. Assuming linear state dynamics, the state constraints can be written as
\begin{equation}
	\un{x}_{k+1}\leq A_k x_k+B_k u_k+C_k p_k\leq\bar{x}_{k+1},\ k=1,\dots,N-1,
	\label{eq:stateDynamics}
\end{equation}
where $u_k$ summarizes all inputs other than power. 

The set of all power trajectories ${p:=[p_1,\dots,p_N]\trans}$ that satisfy constraints \eqref{eq:powerConstr}-\eqref{eq:stateDynamics} is given by the convex polytope, referred to as the resource polytope in \cite{Trangbaek2012}, ${P:=\{p\in\R{N}: Ap\leq b\}}$,
where $A,\,b$ summarize the constraint matrices and the limits of \eqref{eq:powerConstr}-\eqref{eq:stateDynamics}, respectively.

Many energy systems are abstracted as idealized energy buffers subjected to power and energy constraints \eqref{eq:powerConstr} and \eqref{eq:energyConstr} only, see \cite{Gonzalez2015,Zhao2016b}, among others. 
Such systems are referred to as \emph{PE-systems}, and their feasible sets are called \emph{PE-polytopes}.


\subsection{Zonotopic feasible sets}
\label{ss:zonoCharacteristics}
The description of flexibility by polytopes poses serious computational challenges when making decisions for an entire group of systems, \cf\sect\ref{s:aggregation}.
To alleviate this difficulty we propose to inner-approximate individual polytopic feasible sets by a subclass of polytopes, known as zonotopes, that can be aggregated efficiently. 
Zonotopes are centrally symmetric objects. This allows a zonotope $Z$ to be expressed in terms of its center $c\in\R{N}$ and generators ${g^{(i)}\in\R{N}}$, ${\|g^{(i)}\|_2=1}$, ${i=1,\dots,g,}$ as
\begin{equation}
	Z=\{ x\in\R{N}:x=c+G\beta,\ -\bar{\beta}\leq\beta\leq\bar{\beta}\},
	\label{eq:defZonoSym}
\end{equation}
where the generators are summarized in the generator matrix ${G:=[g^{(1)},\dots,g^{(g)}]\in\R{N\times g}}$. The scaling factor ${\beta\in\R{g}}$ is symmetrically bounded by $\bar{\beta}$. The shorthand $Z(G,c,\bar{\beta})$ is used to denote such a zonotope. 
We propose a particular generator matrix such that the associated family of zonotopes is well suited to approximate PE-polytopes. 
For arbitrary dimension ${N\in\mathbb{N}}$, a total of ${2N-1}$ generators are constructed as
\begin{align}
	\gen{i} &:= [0,\dots,0,\overbrace{1}^{i},0,\dots,0]\trans\in\R{N},\ \ts{and}\label{eq:myGen1}\\
	\gen{N+j} &:= [0,\dots,0,\underbrace{-1/\sqrt{2}}_{j},\underbrace{1/\sqrt{2}}_{j+1},0,\dots,0]\trans\in\R{N},
	\label{eq:myGen2}
\end{align}
with ${i=1,\dots,N}$ and ${j=1,\dots,N-1}$.
The corresponding generator matrix is ${G^{PE}\in\R{N\times (2N-1)}}$. Figure \ref{fig:distsZP} provides a two-dimensional example of such a zonotope.
%
It is shown in Appendix \ref{a:zonoNormals} that zonotopes with generator matrix $G^{PE}$ can reconstruct every possible facet of a PE-polytope. However, the approximation quality is influenced by the fact that zonotopes are centrally symmetric by construction, and, in general, will feature additional facets that do not exist in a PE-polytope. 

%% file: zonoApprox.tex
\section{Computing zonotopic approximations}
\label{s:zonoApprox}

\subsection{Computing optimal zonotopes}
\label{ss:computingZono}

\begin{figure}[tb]
	\centering
	\resizebox{0.45\textwidth}{!}{
\begin{pspicture*}(-3,-1.5)(6,4.8)
\psset{unit=0.65cm}

\psline[arrowsize=3pt 3,linewidth=0.8pt]{->}(-2.5,-2)(6,-2)
\psline[arrowsize=3pt 3,linewidth=0.8pt]{->}(-2.25,-2.25)(-2.25,4.5)
\rput[cb](6.4,-2){$p_1$}
\rput[cb](-2.2,4.75){$p_2$}

\pspolygon[fillcolor=lightgray,fillstyle=solid](2,0.25)(4.5,0.25)(4.5,2.25)(3,3.75)(0.5,3.75)(0.5,1.75)
\pspolygon[linewidth=0.8pt](1,0)(5,0)(5,4)(0,4)(0,1)

\psline[linestyle=dotted]{-}(0,1)(0,-1.8)
\psline[linestyle=dotted]{-}(5,1)(5,-1.8)
\psline[arrowsize=3pt 3,linewidth=0.8pt]{<->}(0,-1.6)(5,-1.6)
\rput[cb](2.5,-1.3){$\Delta_{P,1}$}

\psline[linestyle=dotted]{-}(0.5,1.75)(0.5,-1.0)
\psline[linestyle=dotted]{-}(4.5,0.25)(4.5,-1.0)
\psline[arrowsize=3pt 3,linewidth=0.8pt]{<->}(0.5,-0.8)(4.5,-0.8)
\rput[cb](2.5,-0.5){$\Delta_{Z,1}$}

\psline[linestyle=dotted]{-}(0,1)(-2.05,3.05)
\psline[linestyle=dotted]{-}(5,4)(1.95,7.05)
\psline[arrowsize=3pt 3,linewidth=0.8pt]{<->}(-1.9,2.9)(2.1,6.9)
\rput[rb](0.1,5.0){$\Delta_{P,3}$}

\psline[linestyle=dotted]{-}(0.5,1.75)(-1.15,3.4)
\psline[linestyle=dotted]{-}(3,3.75)(1.1,5.65)
\psline[arrowsize=3pt 3,linewidth=0.8pt]{<->}(-1.0,3.25)(1.25,5.5)
\rput[lt](0.5,4.7){$\Delta_{Z,3}$}

\psline[linestyle=dotted]{-}(5,0)(6.65,0)
\psline[linestyle=dotted]{-}(5,4)(6.65,4)
\psline[arrowsize=3pt 3,linewidth=0.8pt]{<->}(6.5,0)(6.5,4)
\rput[cb]{90}(6.2,2){$\Delta_{P,2}$}

\psline[linestyle=dotted]{-}(3,3.75)(5.95,3.75)
\psline[linestyle=dotted]{-}(4.5,0.25)(5.95,0.25)
\psline[arrowsize=3pt 3,linewidth=0.8pt]{<->}(5.65,0.25)(5.65,3.75)
\rput[cb]{90}(5.35,2){$\Delta_{Z,2}$}

\rput(4.4,3.2){$P$}
\rput(1.6,1.5){$Z$}

\psline[arrowsize=3pt 3]{*->}(2.5,2)(3.5,2)
\psline[arrowsize=3pt 3]{->}(2.5,2)(2.5,3)
\psline[arrowsize=3pt 3]{->}(2.5,2)(1.7929,2.7071)

\rput[Bl](3.6,1.85){$\gen{1}$}
\rput[b](2.6,3.0){$\gen{2}$}
\rput[B](1.7,2.7){$\gen{3}$}
\rput[t](2.5,1.85){$c$}

\end{pspicture*}}
	\caption{A PE-polytope $P$ and inscribed zonotope $Z$ whose generators are chosen according to \eqref{eq:myGen1}-\eqref{eq:myGen2}. The distances between parallel facets of the zonotope are referred to as $\Delta_{Z,i}$, ${i=1,\dots,q}$. The extension of the polytope in the direction of the $i^{\ts{th}}$ zonotope facet normal is denoted by $\Delta_{P,i}$.}
	\label{fig:distsZP}
\end{figure}
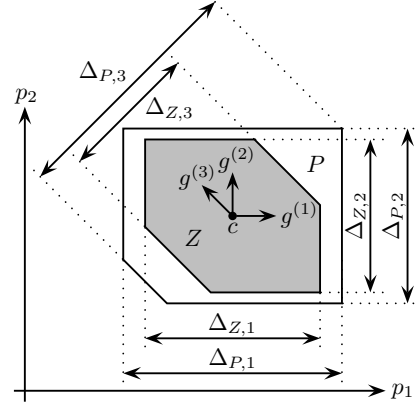

In this section, it is shown how to compute zonotopes that inner-approximate a given polytope and that are optimal with regard to specific objectives. A key ingredient to all the optimization problems formulated below is a set of constraints that enforce that a zonotope is included in a polytope: ${Z(G,c,\bar{\beta})\subseteq P(A,b)\ \Leftrightarrow\ Ac+|AG|\bar{\beta}\leq b}$, \cf\cite{Mueller2015} for more details.
Given a resource polytope $P(A,b)$ and a generator matrix $G$, different objectives can be used to compute zonotopes that inner-approximate the polytope \cite{Mueller2015}. The choice of objective depends on the purpose for which the aggregate feasible set will be used. Here we consider the objective illustrated in \fig\ref{fig:distsZP}. 
The distances between parallel zonotope facets are given as ${\Delta_Z=2|F\trans G|\bar{\beta}}$, where ${F\in\R{N\times q}}$ is a matrix that contains the normal vectors of all possible zonotope facets as columns, \ie, ${F:=[f\ind{1},\dots,f\ind{q}]\trans\in\R{N\times q}}$, with ${||f\ind{i}||_2=1}$, ${i=1,\dots,q}$. Zonotopes with generator matrix $G^{PE}$ can have at most ${q=N^2+N}$ pairs of parallel facets, \cf Appendix \ref{a:zonoNormals}. Similarly, the extension of the polytope in the direction of the $i^{\ts{th}}$ zonotope facet normal is denoted by $\Delta_{P,i}$, ${i=1,\dots,q}$, and computed in a preprocessing step by solving $q$ Linear Programs (LP). Let ${\Delta_P:=[\Delta_{P,1},\dots,\Delta_{P,q}]\trans}$. The zonotope approximates the polytope perfectly if and only if ${\Delta_Z=\Delta_P}$. Thus, a convenient and informative approximation quality measure is
\begin{equation}
	\Lambda := \frac{1}{q}\sum\limits_{i=1}^q{\frac{\Delta_{Z,i}}{\Delta_{P,i}}}\in [0,\,1],
	\label{eq:defApproxQuality}
\end{equation}
where ${\Lambda=0}$ implies that the zonotope is a point, and ${\Lambda=1}$ is reached iff the polytope is approximated perfectly by the zonotope. The inner-approximating zonotope maximizing \eqref{eq:defApproxQuality} can be found by solving the following LP:
\begin{maxi!}
	{\{c,\bar{\beta}\}}{w\trans\bar{\beta} \label{eq:zonoOptiObj}}{\label{eq:zonoOpti}}{}
	\addConstraint{Ac+|AG|\bar{\beta}}{\leq b\label{eq:zonoInclConstr}}
	\addConstraint{\bar{\beta}}{\geq 0, \label{eq:betaConstr}}
\end{maxi!}
with ${w:=(2/q)(1/\Delta_P)\trans|F\trans G|}$. The objective \eqref{eq:zonoOptiObj} expresses \eqref{eq:defApproxQuality} as a linear function of $\bar{\beta}$. Constraints \eqref{eq:zonoInclConstr}-\eqref{eq:betaConstr} guarantee that the zonotope is included in the polytope.

Problem \eqref{eq:zonoOpti} may result in degenerate zonotopes with ${\Delta_{Z,i}=0}$ for some ${i\in \{1,\dots,q\}}$ even though ${\Delta_{P,i}> 0}$. This is undesirable because such zonotopes lack a significant portion of the polytope flexibility. To prevent such degeneracy, \eqref{eq:zonoOpti} can be augmented by constraints enforcing the zonotope to contain a full-dimensional object, \eg, a box. Let $C_{\max}$ denote the maximum-volume axis-aligned cube in $P$. Further, let $B_{\max}$ be the box with maximum  cumulative edge length from the set of axis-aligned boxes ${\{B: C_{\max}\subseteq B\subseteq P\}}$. We augment \eqref{eq:zonoOpti} by convex constraints enforcing that ${B_{\max}\subseteq Z(G,c,\bar{\beta})}$.

\subsection{Approximation results}
\label{ss:approxQuality}
How well a zonotope can approximate a given polytope depends on a number of factors, including the choice of generator matrix, the shape of the polytope (in particular its symmetry properties), and, in case the approximation is based on optimization, the objective function chosen. Moreover, the number of possible facets of the zonotope and the polytope grows with the number of dimensions $N$ and also affects the approximation quality. 	
Here, the approximation quality of zonotopes is assessed for resource polytopes of plug-in electric vehicles (PEVs) over a planning horizon of 24 h with 2 h sampling time. It is assumed that the PEVs are subject to power and energy constraints only. Trip constraints can easily be translated into power constraints by setting both the upper and lower power bounds to zero for all time slots that fall into trip periods. Table \ref{tab:sysParams} provides the parameter ranges from which the parameters of 100 PEVs are sampled uniformly. For every system, an optimal inner-approximating axis-aligned box $B\opt$ and zonotope $Z\opt$ are computed via \eqref{eq:zonoOpti} using ${G=I_N}$ and ${G=G^{PE}}$, respectively. 
%
%
%
Averaged over the population, the quality of approximation \eqref{eq:defApproxQuality} for the box and the zonotope are ${\Lambda_B=0.31}$ and ${\Lambda_Z=0.63}$, respectively.

\begin{table}[b]
\setlength{\abovecaptionskip}{-5pt}
\caption{PEV parameter ranges.}
\label{tab:sysParams}
\begin{center}
\begin{tabular}{|c|l|c|} \hline
Variable & Description & Value range\\ \hline 
$H$ & Planning horizon & 24 h\\
$\bar{p}$ & Maximum rated power & 3 kW\\
$\un{p}$ & Minimum rated power & -3 kW\\
$\bar{e}-\un{e}$ & Battery capacity & [20,40] kWh \\
$(e_0-\un{e})/(\bar{e}-\un{e})$ & Initial state-of-charge & [0.2,0.8]\\
\hline
\end{tabular}
\end{center}
\end{table}

%


%% file: aggregation.tex
\section{Aggregation of flexibility}
\label{s:aggregation}




\subsection{Aggregation setup}
To take full advantage of the flexibility available from entire groups of flexible energy resources, their individual flexibilities are pooled by aggregators. Aggregators act as intermediaries between flexible systems and wholesale markets. Consider an aggregator that manages the electric energy consumption/production of a population of systems indexed by ${j\in\mathcal{J}:=\{1,\dots,J\}}$. Every system provides the aggregator with a description of its feasible set ${\fset\ind{j}\subseteq\R{N}}$ for a given planning horizon comprising $N$ time steps. The feasible set of a system is either known directly from the physical properties and constraints of the system or has to be estimated from measurement data \cite{Mueller2017}. 
%
%
The aggregator decides how to optimally use the flexibility available from all the systems in the population. The aggregator might trade electricity on wholesale energy markets or offer dedicated services to flexibility markets, \eg, regulation power. 
Finally, the aggregator is required to assign to every system a reference schedule ${p\ind{j}\in\fset\ind{j}}$, ${j\in\mathcal{J}}$, that the system is committed to follow. This assignment is done either ahead of time, \eg, in the case of a day-ahead reference schedule, or during time of delivery, \eg, upon the activation of regulating power. Given the electricity consumption/production of an individual system during the period of delivery, money flows either from the aggregator to the system or vice versa according to the terms agreed upon by the system and the aggregator.

\subsection{Aggregation of feasible sets}
\label{ss:aggFlex}
A power trajectory ${p\ind{\ts{agg}}\in\R{N}}$ that can be followed collectively by the population of $J$ systems is called \emph{aggregate feasible}. 
The set of aggregate feasible power trajectories is denoted by $\fset\ind{\ts{agg}}$ and referred to as the \emph{aggregate feasible set}. It is given by the Minkowski sum of all individual feasible sets \cite{Hao2013,Barot2016}:
\begin{align}
	\fset\ind{\ts{agg}} :=& \,\fset\ind{1}\oplus\dots\oplus \fset\ind{J}\nonumber\\
	 =& \Bigg\{p\in\R{N}: p=\sum\limits_{j\in \mathcal{J}}p^{(j)},\,p^{(j)}\in \fset^{(j)}\Bigg\}.
	 \label{eq:aggFeasSet}
\end{align}

The next sections look into computing $\fset\ind{\ts{agg}}$ for the special cases where the individual feasible sets are given either by general polytopes or by zonotopes.

\textit{1) Aggregation of polytopic flexibility:} When the feasible sets $\fset\ind{j}$ are given by polytopes the aggregate feasible set \eqref{eq:aggFeasSet} is also a polytope \cite{Weibel2007}. Different methods to compute the Minkowski sum of polytopes exist, for example by computing the convex hull of the pairwise sum of all vertices of the polytopes or by projecting an augmented polytope to a lower-dimensional subspace \cite{Kvasnica2005,Weibel2007,Zhao2016b}. However, no efficient algorithm exists to compute the Minkowski sum of general polytopes in hyper-plane representation \cite{Kvasnica2005,Weibel2007}. Consequently, attempts to compute the Minkowski sum of resource polytopes are restricted to special types of polytopes (symmetric, binary hyperplane normal vectors) \cite{Hao2013,Trangbaek2011}, are limited to low dimensions $N$, or compute approximations of the true sum \cite{Zhao2016b,Alizadeh2014}.

\textit{2) Aggregation of zonotopic flexibility:} The Minkowski sum of zonotopes can be computed efficiently. The aggregate feasible set $Z\ind{\ts{agg}}$ of the individual feasible sets $Z(G,c\ind{j},\bar{\beta}\ind{j})$ is given explicitly by
\begin{align}
\begin{split}
	Z\ind{\ts{agg}}:=\, & Z(G,c\ind{\ts{agg}},\bar{\beta}\ind{\ts{agg}})\\
	 =\,& Z\bigg(G,\sum\limits_{j\in\mathcal{J}}{c\ind{j}},\sum\limits_{j\in\mathcal{J}}{\bar{\beta}\ind{j}}\bigg).
	 \end{split}
	 \label{eq:defAggZono}
\end{align}
The structure of \eqref{eq:defAggZono} is particularly convenient for aggregators. First, $Z\ind{\ts{agg}}$ is a zonotope because zonotopes are closed under Minkowski addition. Second, if systems are added to or removed from the population or if individual feasible sets change, the aggregate feasible set can be adjusted efficiently by updating the sums $c\ind{\ts{agg}}$ and $\bar{\beta}\ind{\ts{agg}}$ accordingly.

%% file: disaggregation.tex
\section{Disaggregation}
\label{s:disaggregation}
%
%
%
%
%
\subsection{Problem formulation}
\label{ss:disaggProblem}
In the aggregation scheme introduced above, the aggregator is responsible for trading a sufficient amount of electric energy such that every system can follow a feasible power trajectory ${p\ind{j}\in\mathcal{P}\ind{j}}$, ${j\in\mathcal{J}}$. The contract between the aggregator and each individual system $j$ defines the price $v_k\ind{j}$ at which the system buys/sells electric energy from/to the aggregator for every time step ${k=1,\dots,N}$. The aggregator assigns a reference trajectory $p\ind{j}$ to every system, for which the system pays the amount ${t_s v^{(j)^{\top}}p\ind{j}}$, ${v\ind{j}:=[v\ind{j}_1,\dots,v\ind{j}_N]\trans}$, to the aggregator. While the systems guarantee that every ${p\ind{j}\in\mathcal{P}\ind{j}}$ can be implemented, they might prefer certain trajectories over others. For example, the owner of a building might prefer those power trajectories that keep the indoor air-temperature within certain limits. Similarly, the operator of a gas power plant might favor constant power trajectories over ones that vary quickly because of ramping costs. 
To express particular preferences and additional operating costs, every system defines a convex \emph{flexibility cost function} ${f\ind{j}:\mathcal{P}\ind{j}\rightarrow \R{}}$.
If we assume that the aggregator reimburses the individual systems for their flexibility costs, the total costs $T\ind{j}(p\ind{j})$ of having system $j$ implementing the power trajectory $p\ind{j}$ are given by the difference of the flexibility costs and the energy costs: 
\begin{equation}
	T\ind{j}(p\ind{j}):=f\ind{j}(p\ind{j})-t_s v^{(j)^\top} p\ind{j}. 
	\label{eq:trackingCost}
\end{equation}
If $T\ind{j}$ is negative, money flows from the system owner to the aggregator. If, in contrast, the costs are positive, the system owner receives money from the aggregator.
%
%

The disaggregation problem the aggregator needs to solve is how to best distribute a given trajectory ${p\ind{\ts{agg}}\in\fset\ind{\ts{agg}}}$ into exactly $J$ trajectories ${p\ind{1},\dots,p\ind{J}}$ such that they are feasible and sum up to $p\ind{\ts{agg}}$. This disaggregation is not unique in general, and the aggregator may want to find the one that is optimal with regard to a particular objective. A profit-oriented aggregator will  minimize the total costs by solving
\begin{mini}
	{\{p\ind{j}\}_{j\in\mathcal{J}}}{\sum\limits_{j\in\mathcal{J}} T\ind{j}(p\ind{j})}
	{\label{eq:disagOrig}}{}
	\addConstraint{\sum\limits_{j\in\mathcal{J}} p\ind{j}}{= p\ind{\ts{agg}}}
	\addConstraint{p\ind{j}}{\in\fset\ind{j},\ j\in\mathcal{J}.}
\end{mini}

\subsection{Disaggregation using zonotopes}
\label{ss:computeAggCost}
\begin{figure}[tbh]
\setlength{\abovecaptionskip}{-2pt}
\centering
\resizebox{0.45\textwidth}{!}{
\begin{pspicture*}(-3.75,-2.2)(4.5,6)
	\psline[linestyle=dotted,linecolor=gray]{}(-2.5,5.0)(-2.5,2.75)
	\psline[linestyle=dotted,linecolor=gray]{}(-1.75,4.25)(-1.75,2.75)
	\psline[linestyle=dotted,linecolor=gray]{}(-0.75,5.0)(-0.75,2.75)
	
	\psaxes[Dx=1]{->}(-1.75,4.0)(-3,4.0)(-0.4,5.5)
	\psline[linewidth=1.0pt,showpoints=true]{}(-2.5,4.25)(-1.75,4.25)(-0.75,4.5)
	\psline[linewidth=0.8pt]{<->}(-2.5,3)(-1.75,3)
	\psline[linewidth=0.8pt]{<->}(-1.75,3)(-0.75,3)

	\rput[l](-0.35,4){$\beta\ind{1}_i$}
	\rput[lb](-1.85,5.5){$T\ind{1}_i(\beta\ind{1}_i)$}
	
	\rput[b](-2.1,3.05){$l\ind{1}_{i,1}$}
	\rput[b](-1.25,3.05){$l\ind{1}_{i,2}$}
	\rput[b](-2.1,4.3){$q\ind{1}_{i,1}$}
	\rput[b](-1.25,4.5){$q\ind{1}_{i,2}$}
	
	\psline[linestyle=dotted,linecolor=gray]{}(1.15,5.0)(1.15,2.75)
	\psline[linestyle=dotted,linecolor=gray]{}(2.15,4.25)(2.15,2.75)
	\psline[linestyle=dotted,linecolor=gray]{}(3.15,5.0)(3.15,2.75)
	
	\psaxes[Dx=1]{->}(2.15,4)(0.9,4)(3.5,5.5)
	\psline[linewidth=1.0pt,showpoints=true]{}(1.15,4.5)(2.15,4)(3.15,4.75)
	\psline[linewidth=0.8pt]{<->}(1.15,3)(2.15,3)
	\psline[linewidth=0.8pt]{<->}(2.15,3)(3.15,3)
	
	\rput[l](3.55,4){$\beta\ind{2}_i$}
	\rput[lb](2.05,5.5){$T\ind{2}_i(\beta\ind{2}_i)$}
	
	\rput[b](1.7,3.05){$l\ind{2}_{i,1}$}
	\rput[b](2.7,3.05){$l\ind{2}_{i,2}$}
	\rput[b](1.7,4.3){$q\ind{2}_{i,1}$}
	\rput[b](2.6,4.5){$q\ind{2}_{i,2}$}
	
	\psline[linestyle=dotted,linecolor=gray]{}(-1.75,1.75)(-1.75,-1.5)
	\psline[linestyle=dotted,linecolor=gray]{}(-0.75,1.75)(-0.75,-1.5)
	\psline[linestyle=dotted,linecolor=gray]{}(0,0)(0,-1.5)
	\psline[linestyle=dotted,linecolor=gray]{}(1.0,1.75)(1.0,-1.5)
	\psline[linestyle=dotted,linecolor=gray]{}(2.0,1.75)(2.0,-1.5)
	
	\psaxes[Dx=1]{->}(0,0)(-2.25,0)(2.5,2.3)
	\psline[linewidth=1.0pt,showpoints=true]{}(-1.75,0.75)(-0.75,0.25)(0,0.25)(1,0.5)(2,1.25)
	\psline[linewidth=0.8pt]{<->}(-1.75,-1.2)(-0.75,-1.2)
	\psline[linewidth=0.8pt]{<->}(-0.75,-1.2)(0,-1.2)
	\psline[linewidth=0.8pt]{<->}(0,-1.2)(1,-1.2)
	\psline[linewidth=0.8pt]{<->}(1,-1.2)(2,-1.2)
	
	\rput[l](2.55,0){$\beta\ind{\ts{agg}}_i$}
	\rput[lb](0.15,2.1){$T\ind{\ts{agg}}_i(\beta\ind{\ts{agg}}_i)$}
	
	\rput[b](-1.2,-1.2){$l\ind{\ts{agg}}_{i,1}$}
	\rput[b](-0.35,-1.2){$l\ind{\ts{agg}}_{i,2}$}
	\rput[b](0.55,-1.2){$l\ind{\ts{agg}}_{i,3}$}
	\rput[b](1.55,-1.2){$l\ind{\ts{agg}}_{i,4}$}
	\rput[b](-1.2,0.65){$q\ind{\ts{agg}}_{i,1}$}
	\rput[b](-0.4,0.3){$q\ind{\ts{agg}}_{i,2}$}
	\rput[b](0.55,0.45){$q\ind{\ts{agg}}_{i,3}$}
	\rput[b](1.5,1.0){$q\ind{\ts{agg}}_{i,4}$}
	
	\rput[c](-1.6,-1.8){$-\bar{\beta}\ind{\ts{agg}}_{i}$}	
	\rput[c](2.3,-1.8){$\bar{\beta}\ind{\ts{agg}}_i$}	
	
\end{pspicture*}}
\caption{Examples of cost function components, $T\ind{1}_i(\beta\ind{1}_i)$ and $T\ind{2}_i(\beta\ind{2}_i)$ (top). Aggregate cost function component $T\ind{\ts{agg}}_i(\beta\ind{\ts{agg}}_i)$ derived from $T\ind{1}_i$ and $T\ind{2}_i$ by concatenating line segments ordered by ascending slopes (bottom).}
\label{fig:pwlFun}
\end{figure}
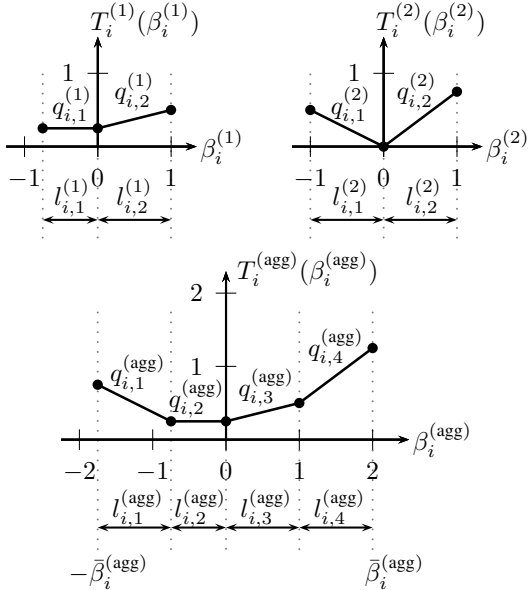

We consider feasible sets $\fset\ind{j}$ given by the zonotopes ${\{p\ind{j}\in\R{N}: p\ind{j}=c\ind{j}+G\beta,\,-\bar{\beta}\ind{j}\leq\beta\leq\bar{\beta}\ind{j}\}}$, ${j\in\mathcal{J}}$. The particular structure of the Minkowski sum of zonotopes \eqref{eq:defAggZono} allows an efficient disaggregation algorithm whose complexity scales particularly well with the number of systems in the population. To exploit this advantage, we restrict attention to flexibility costs that can be expressed as a function of $\beta$. In particular, we consider cost functions ${f\ind{j}(\beta\ind{j}): \R{g}\rightarrow\R{}}$ that can be written as ${f\ind{j}(\beta\ind{j}) = f\ind{j}_1(\beta\ind{j}_1)+\dots+f\ind{j}_g(\beta\ind{j}_g)}$, 
where each ${f\ind{j}_i:\R{}\rightarrow\R{}}$ is piece-wise linear convex, and $\beta_i\ind{j}$ refers to the $i^{\ts{th}}$-element of the vector ${\beta\ind{j}\in\R{g}}$. Substituting $f\ind{j}(p\ind{j})$ with $f\ind{j}(\beta\ind{j})$ and $p\ind{j}$ with ${c\ind{j}+G\beta\ind{j}}$ allows to rewrite the total costs \eqref{eq:trackingCost} in terms of $\beta$ as
\begin{align}
	T\ind{j}(\beta\ind{j}) 
	&= \sum\limits_{i=1}^g\left(f\ind{j}_i(\beta\ind{j}_i) - t_s v^{(j)\trans}g\ind{i}\beta\ind{j}_i\right)+T\ind{j}_{\ts{fix}}\nonumber\\
	&= \sum\limits_{i=1}^g{T\ind{j}_i(\beta\ind{j}_i)}+T\ind{j}_{\ts{fix}}\label{eq:indTotCost},
\end{align} 
where $g\ind{i}$ is the $i^{\ts{th}}$ generator, and ${T\ind{j}_{\ts{fix}}:=-t_sv^{(j)\trans}c\ind{j}}$ is a fixed cost independent of $\beta\ind{j}$. The components $T_i\ind{j}(\beta_i\ind{j})$ are piece-wise linear convex. For the purpose of illustration, \fig\ref{fig:pwlFun} shows two example cost functions $T_i\ind{1}$ and $T_i\ind{2}$. The breakpoint interval lengths and the slopes of the line segments are denoted by $l\ind{j}_{i,t}$ and $q\ind{j}_{i,t}$, respectively, for ${t=1,\dots,m\ind{j}_i}$ with $m\ind{j}_i$ being the number of line segments.

%

We define the \emph{aggregate} cost $T\ind{\ts{agg}}(\beta\ind{\ts{agg}})$ as the total cost of the cheapest disaggregation of $\beta\ind{\ts{agg}}$ into ${\beta\ind{1},\dots,\beta\ind{J}}$:
\begin{mini}[3]
{\{\beta\ind{j}\}_{j\in\mathcal{J}}}
{\sum\limits_{j\in\mathcal{J}}T\ind{j}(\beta\ind{j})}
{\label{eq:defAggTrackCost}}
{T\ind{\ts{agg}}(\beta\ind{\ts{agg}}):=}
\addConstraint{\sum\limits_{j\in\mathcal{J}}\beta\ind{j}}{= \beta\ind{\ts{agg}}}
\addConstraint{-\bar{\beta}\ind{j}}{\leq\beta\ind{j}\leq\bar{\beta}\ind{j},\ j\in\mathcal{J}.}
\end{mini}
%
%
The structure of the individual costs \eqref{eq:indTotCost} implies that $T\ind{\ts{agg}}$ can be written as
\begin{equation}
	T\ind{\ts{agg}}(\beta\ind{\ts{agg}})=\sum\limits_{i=1}^g{T_i\ind{\ts{agg}}(\beta_i\ind{\ts{agg}})}+T_{\ts{fix}}\ind{\ts{agg}},
\end{equation}
with ${T_{\ts{fix}}\ind{\ts{agg}}:=T_{\ts{fix}}\ind{1}+\dots+T_{\ts{fix}}\ind{J}}$. Each component $T\ind{\ts{agg}}_i$ is piece-wise linear convex and can be constructed by concatenating the line segments of ${T_i\ind{1},\dots,T_i\ind{J}}$ sorted by ascending slopes $q_{i,t}\ind{j}$, see \fig\ref{fig:pwlFun}. The offset of $T_i\ind{\ts{agg}}$ is given by 
\begin{equation*}
	T\ind{\ts{agg}}_i(\bar{\beta}\ind{\ts{agg}}_i)=\sum\limits_{j\in\mathcal{J}} T\ind{j}_i(\bar{\beta}\ind{j}_i)+T_{\ts{fix}}\ind{\ts{agg}}.
\end{equation*}
The variable $s_{i,t}$ stores the system index that belongs to line segment $t$ of $T_i\ind{\ts{agg}}$: ${s_{i,1}=2}$ for the example in \fig\ref{fig:pwlFun}.  Up to its offset, each cost component $T_i\ind{\ts{agg}}$, ${i=1,\dots,g}$, is fully defined by the ordered list of triples ${L_i:=\{(l_{i,1}\ind{\ts{agg}},q_{i,1}\ind{\ts{agg}},s_{i,1}),\dots,(l_{i,M_i}\ind{\ts{agg}},q_{i,M_i}\ind{\ts{agg}},s_{i,M_i})\}}$, with ${M_i:=m_i\ind{1}+\dots+m_i\ind{J}}$.
%
These lists are key prerequisites for the disaggregation algorithm discussed below.

\subsection{Subgradient-based disaggregation algorithm}
\label{ss:disagZono_sg}
Given an aggregate trajectory ${p\ind{\ts{agg}}\in Z\ind{\ts{agg}}}$, the feasible sets $Z(G,c\ind{j},\bar{\beta}\ind{j})$, ${j\in\mathcal{J}}$, and the aggregate cost function $T\ind{\ts{agg}}(\beta\ind{\ts{agg}})$, the disaggregation problem is to solve
\begin{mini}
{\beta}
{T\ind{\ts{agg}}(\beta)}
{\label{eq:disagOptiSG}}{}
\addConstraint{p\ind{\ts{agg}}}{= c\ind{\ts{agg}}+ G \beta}
\addConstraint{-\bar{\beta}\ind{\ts{agg}}}{\leq\beta\leq\bar{\beta}\ind{\ts{agg}}.}
\end{mini}
The key difference between the above problem and its original formulation \eqref{eq:disagOrig} is that in \eqref{eq:disagOptiSG} the number of decision variables ($g$), the number of equality constraints ($N$), and the number of inequality constraints ($2g$) are independent of the population size $J$.

Problem \eqref{eq:disagOptiSG} is convex with piece-wise linear objective and can be written as an LP using an epigraph reformulation, with the objective substituted by a set of linear inequalities. This is inconvenient for the application discussed here because the number of inequalities can be large. Instead, problem \eqref{eq:disagOptiSG} is solved via a \emph{projected subgradient method} \cite{Boyd2003}. 
%
%
The disaggregation algorithm proposed here comprises four steps:
\begin{enumerate}[noitemsep]
	\item \textit{Initialization:} Find $\beta(0)$ in the feasible set of \eqref{eq:disagOptiSG}.
	\item \textit{Subgradient steps:} In iteration $k$, compute a subgradient of $T\ind{\ts{agg}}$ at $\beta(k)$ and execute a subgradient step. 
	\item \textit{Projection:} Project the new ${\beta(k+1)}$ back onto the feasible set of \eqref{eq:disagOptiSG}. Repeat 2) and 3) until converged, and denote the best $\beta$ found so far by $\beta^{(\ts{agg})\ast}$.
	\item \textit{Assignment:} Distribute $\beta^{(\ts{agg})\ast}$ into $\beta^{(1)\ast},\dots,\beta^{(J)\ast}$.
\end{enumerate}

The next sections elaborate on the above steps.

\textit{1) Initialization:} The subgradient method requires a feasible starting point ${\beta(0)\in\R{g}}$. It can be obtained by projecting an arbitrary ${\beta\in\R{g}}$ onto the feasible set of \eqref{eq:disagOptiSG}.
\textit{2) Subgradient step:} The subgradient method iteratively executes a subgradient step that updates the decision variable as ${\beta(k+1) = \beta(k)-\alpha(k)\nabla(k)}$, 
where the subgradient and the step size of iteration $k$ are denoted by $\nabla(k)$ and $\alpha(k)$, respectively. A subgradient of $T\ind{\ts{agg}}$ at $\beta(k)$ is computed element-wise according to line 6 in \fig\ref{alg:sgAlg}. 
%
%
Several convergence results for the projected sugradient method exist in the literature \cite{Boyd2003}, including different step size rules that influence the convergence properties. Here, ${\alpha(k)=a/k}$, ${a>0}$, is used, for which the subgradient method converges to the optimal value, \ie, ${\lim_{k\rightarrow\infty} T(\beta(k))=T\opt}$ \cite{Boyd2003}. Setting ${a=(\bar{\beta}\ind{\ts{agg}}_1+\dots+\bar{\beta}\ind{\ts{agg}}_g)/g}$ proved to be effective.

\textit{3) Projection:} The $\beta$ updated by the subgradient step in general does not satisfy all constraints of \eqref{eq:disagOptiSG}. To make it feasible, it is projected onto to the feasible set using the Euclidean projection $Q(\cdot)$, see line 9 in \fig\ref{alg:sgAlg}.

The subgradient method is not a descent method. Thus, at every iteration, the best objective value found so far is stored. The subgradient algorithm is stopped once the termination criterion in line 15 in \fig\ref{alg:sgAlg} is met. The optimal $\beta^{(\ts{agg})\ast}$ and the corresponding objective value $V(k\opt)$ are returned.
%

\begin{figure}[h!]
\begin{boxedalgorithmic}[1]
\REQUIRE $p\ind{\ts{agg}},\,c\ind{\ts{agg}},\,\beta(0),\,\bar{\beta}\ind{\ts{agg}},\,L_1,\dots,L_g,$ $T_{\ts{fix}}\ind{\ts{agg}},\,a,\,h,\,\epsilon$
\STATE $V(0)\leftarrow \ts{INF}$, $k\leftarrow 0$
\REPEAT
\STATE $W\leftarrow T_{\ts{fix}}\ind{\ts{agg}}$

\FOR{$i=1$ to $g$}
\STATE Find largest $m:\sum\limits_{t=1}^{m}{l\ind{\ts{agg}}_{i,t}}\leq\bar{\beta}\ind{\ts{agg}}_i+\beta_i(k)$
\STATE $\nabla_i(k) \leftarrow q\ind{\ts{agg}}_{i,\min(m+1,M_i)}$
\STATE ${W\leftarrow W+\sum\limits_{t=1}^{m}{l\ind{\ts{agg}}_{i,t} q\ind{\ts{agg}}_{i,t}}}$
 $+\left(\bar{\beta}\ind{\ts{agg}}_i+\beta_i(k)-\sum\limits_{t=1}^m{l\ind{\ts{agg}}_{i,t}}\right)q\ind{\ts{agg}}_{i,\min(m+1,M_i)}$
\ENDFOR

\STATE $\beta(k+1)\leftarrow Q(\beta(k)-a/(k+1)\nabla(k))$

\STATE $V(k+1)\leftarrow\min(V(k), W)$

\IF{$V(k+1)<V(k)$}
\STATE $k\opt\leftarrow k+1$
\STATE $\beta^{(\ts{agg})\ast}\leftarrow\beta(k\opt)$
\ENDIF

\STATE $k \leftarrow k+1$
\UNTIL{${(V(\max(0,k-h))-V(k))/V(k)}\leq\epsilon$}
\RETURN $\beta^{(\ts{agg})\ast},\,V(k\opt)$
\end{boxedalgorithmic}
\caption{Projected subgradient algorithm.}
\label{alg:sgAlg}
\end{figure}

\textit{4) Distribution to individual values:} Once the subgradient method has terminated, the best choice of the decision variable $\beta^{(\ts{agg})\ast}$ has to be distributed to individual ${\beta^{(1)\ast},\dots,\beta^{(J)\ast}}$. This is done by the algorithm presented in \fig\ref{alg:disag}. Finally, the corresponding individual minimum-cost power trajectories are given by ${p^{(j)\ast} = c\ind{j}+G\beta^{(j)\ast},\ j\in\mathcal{J}}$.

\begin{figure}[h!]
\begin{boxedalgorithmic}[1]
\REQUIRE $\beta^{(\ts{agg})\ast}$, $\bar{\beta}\ind{\ts{agg}},\,L_1,\dots,L_g$
\STATE $\beta^{(j)\ast}\leftarrow\mathbf{0}_{g\times 1},\ \forall j\in\mathcal{J}$
\FOR{$i=1$ to $g$}
\STATE Find largest $m:\ \sum\limits_{t=1}^{m}{l_{i,t}\ind{\ts{agg}}}\leq\bar{\beta}_i\ind{\ts{agg}}+\beta^{(\ts{agg})\ast}_i$
\IF{$m>0$}
\STATE $\beta^{(s_{i,t})\ast}_i\leftarrow\bar{\beta}\ind{s_{i,t}}_i,\quad t=1,\dots,m$
\ENDIF
\STATE $\beta^{(s_{i,min(m+1,M_i)})\ast}_i\leftarrow\bar{\beta}_i\ind{\ts{agg}}+ \beta^{(\ts{agg})\ast}_i-\sum\limits_{t=1}^{m}{l\ind{\ts{agg}}_{i,t}}$

\ENDFOR
\RETURN $\beta^{(1)\ast},\dots,\beta^{(J)\ast}$
\end{boxedalgorithmic}
\caption{Distribution of $\beta^{(\ts{agg})\ast}$ into $\beta^{(1)\ast},\dots,\beta^{(J)\ast}$.}
\label{alg:disag}
\end{figure}

%

\subsection{Performance of subgradient-based disaggregation}
To assess the performance of the subgradient-based disaggregation, its solving time is compared with the solving time of the original LP formulation \eqref{eq:disagOrig} for different populations of PEVs sampled uniformly from the parameter ranges provided in \tab\ref{tab:sysParams}. The flexibility costs $f\ind{j}$ are assumed linear with random costs. The optimization models and the subgradient algorithm are implemented in \textsc{Matlab}\textregistered. Optimization problems are solved by \textsc{IBM ILOG CPLEX}\textregistered\footnote{\label{fnote}IBM, ILOG, and CPLEX are trademarks of International Business Machines Corp., registered in many jurisdictions worldwide. Other product and service names might be trademarks of IBM or other companies.} v.12.6  run on a desktop PC featuring an Intel\textregistered\, Core i5-2400 CPU @ 3.1 GHz and 16 GB of RAM.

The original disaggregation problem \eqref{eq:disagOrig} is convex and can, in theory, be solved efficiently. However, the problem grows with the population size $J$ and the number of time steps $N$. In the case of PE-systems, problem \eqref{eq:disagOrig} comprises $NJ$ decision variables, $N$ equalities, and ${J(4N-2)}$ inequalities. Therefore, even in the simplest case of linear costs, the solving time and memory requirements of \eqref{eq:disagOrig} can be significant. Table \ref{tab:disagSolTime} summarizes the solving time for different problem sizes. For some combinations of $N$ and $J$, the problem size exceeds the maximum memory that \textsc{Matlab} can allocate (abbreviated by M). We note that memory is exhausted for problem sizes of practical interest; for example, participation in the weekly secondary reserve market with 15 min intervals would require the solution of a problem with ${N=672}$. 

\begin{table}[tbh]
\setlength{\abovecaptionskip}{-5pt}
\caption{Solving time of LP-based disaggregation \eqref{eq:disagOrig}.}
\label{tab:disagSolTime}
\begin{center}
\begin{tabular}{|r|c|c|c|c|c|c|}  \hline
   {\centering $J$ = }   &  $N$ = 24 &     48 &      96 &      168 & 288 & 672\\ \hline 
${10^3}$  &  2.9 s &  15.9 s &   34.0 s &    88.9 s & 871.7 s  & M\\
${2\cdot 10^3}$  &  5.9 s &  45.7 s &  145.4 s &   670.5 s & 1603.7 s & M\\
${5\cdot 10^3}$  & 26.3 s & 155.9 s &  931.5 s &  2647.4 s & M & M\\
$ 10^4$ & 99.8 s & 449.9 s & 3811.1 s & 7319.2 s & M & M\\ \hline
\end{tabular}
\end{center}
\end{table}

The solving time of the subgradient-based disaggregation depends on the termination parameters $h$ and $\epsilon$. The values ${h=5}$ and ${\epsilon=10^{-3}}$ have proved to yield good results in practice: the projected subgradient method terminates with a relative optimality gap of less than ${2.5\cdot 10^{-3}}$ for all experiments executed. The maximum solving time for different choices of $J$ and $N$ are provided in \tab\ref{tab:solTimeSG}. The subgradient-based disaggregation clearly outperforms the LP-based disaggregation \eqref{eq:disagOrig}. Because the number of decision variables and constraints are independent of $J$, the problem can be solved for large populations and long planning horizons.

\begin{table}[tbh]
\setlength{\abovecaptionskip}{-5pt}
\caption{Solving time of subgradient-based disaggregation.}
\label{tab:solTimeSG}
\begin{center}
\begin{tabular}{|r|c|c|c|c|c|c|}  \hline
    {\centering $J$ = }  &  $N$=24 &    48 &    96 &      168 & 288 & 672\\ \hline 
${10^3}$ & 0.13 s & 0.13 s & 0.17 s & 0.24 s & 0.32 s & 0.70 s\\
  ${2\cdot 10^3}$ & 0.12 s & 0.15 s & 0.22 s & 0.29 s & 0.42 s & 0.88 s\\
  ${5\cdot 10^3}$ & 0.15 s & 0.20 s & 0.27 s & 0.47 s & 0.76 s & 1.71 s\\
  ${10^4}$ & 0.19 s & 0.29 s & 0.48 s & 0.82 s & 1.30 s & 2.97 s\\ \hline
\end{tabular}
\end{center}
\end{table}

%% file: zonoUseCases.tex
\section{Use-cases for zonotopic flexibility}
\label{s:zonoUseCases}



\subsection{Feasibility}
Checking the feasibility of an aggregate trajectory $p\ind{\ts{agg}}$ for polytopic feasible sets requires solving the disaggregation problem \eqref{eq:disagOrig}, which can be computationally expensive. In contrast, checking the feasibility in the case of zonotopes is easy because the aggregate feasible set can be computed explicitly via \eqref{eq:defAggZono}. It holds that ${p\ind{\ts{agg}}\in Z\ind{\ts{agg}}}$ if and only if ${|F\trans(p\ind{\ts{agg}}-c\ind{\ts{agg}})|\leq|F\trans G|\bar{\beta}\ind{\ts{agg}}}$, 
where ${F\in\R{N\times q}}$ is the matrix with the normal vectors of all possible zonotope facets as columns, \cf Appendix \eqref{a:zonoNormals} and \cite{Althoff2010}. That is, checking whether a given trajectory is aggregate feasible amounts to validating a set of $q$ inequalities. 
\subsection{The cost of offering regulation power}
Aggregators can use flexibility to offer ancillary services. Here, a symmetric constant regulation power service is considered. Offering $\bar{r}$ units of regulation power requires that in every time step of the planning horizon, a deviation of $\bar{r}$ from the baseline power $p\ind{\ts{agg}}$ is possible. In the power space, this translates into reserving an axis-aligned cube with edge length $2\bar{r}$ and center $p\ind{\ts{agg}}$ in the aggregate feasible set. A lower bound $\bar{r}_{\max}$ on the maximum regulation power available from the population of systems is
%
%
\begin{maxi!}
{\bar{r},\{p\ind{j},\,r\ind{j}\}_{j\in\mathcal{J}}}{\bar{r}}
{\label{eq:maxCap}}{}
\addConstraint{B(p\ind{j},r\ind{j})}{\subseteq\mathcal{P}\ind{j}, j\in\mathcal{J} \label{eq:boxInclConstr}}
\addConstraint{r\ind{j}}{\geq 0,\ j\in\mathcal{J}}
\addConstraint{\sum\limits_{j\in\mathcal{J}}r\ind{j}}{\geq \bar{r}\mathbf{1} \label{eq:capConstr}}
\end{maxi!}
with ${B(p\ind{j},r\ind{j}):=\{p\in\R{N}: p\ind{j}-r\ind{j}\leq p\leq p\ind{j}+r\ind{j}\}}$.
%
For a fixed amount of regulation power ${\bar{r}\in[0,\bar{r}_{\max}]}$ and an expected energy wholesale price $\hat{v}$, the aggregator can find the cheapest baseline by solving
\begin{equation}
	\hat{R}(\bar{r}):= \min\limits_{\{p\ind{j},r\ind{j}\}_{j\in\mathcal{J}}}  \sum\limits_{j\in\mathcal{J}}\left\{ T\ind{j}(p\ind{j})+\hat{v}\trans p\ind{j} \right\},
		\label{eq:defBaseCosts}
\end{equation} 
subject to constraints \eqref{eq:boxInclConstr}-\eqref{eq:capConstr}.
The expected cost of offering $\bar{r}$ units of regulation power is ${\hat{R}(\bar{r})-\hat{R}(0)}$. This quantity can serve as a lower bound on the regulation power bid price.

However, problem \eqref{eq:defBaseCosts} with polytopic feasible sets has the same complexity as the disaggregation problem \eqref{eq:disagOrig} and thus is inconvenient for large populations and long planning horizons. In contrast, for zonotopic feasible sets, an approximative solution to \eqref{eq:defBaseCosts} can be found efficiently by using the subgradient-based disaggregation algorithm of \sect\ref{s:disaggregation}.
In a first step, $\bar{r}_{\max}$ is computed by solving \eqref{eq:maxCap} reformulated for zonotopes with generator matrices $G^{PE}$ as
\begin{equation}
\begin{aligned}
\bar{r}_{\max}=& \argmax\limits_{\{\beta\ind{j},\bar{r}\}_{j\in\mathcal{J}}} & & \bar{r}\\
			& \hspace*{3mm} \ts{s.t.} & & G^{PE}\beta\ind{j}\geq \mathbf{0},\ j\in\mathcal{J},\\
			& & &G^{PE}\sum\limits_{j\in\mathcal{J}}\beta\ind{j}\geq \bar{r}\mathbf{1},\\
			& & &-\bar{\beta}\ind{j}\leq\beta\ind{j}\leq\bar{\beta}\ind{j},\ j\in\mathcal{J}.
		\end{aligned}
		\label{eq:maxCapBetas}
\end{equation}
Denote by ${\beta\ind{j}_{\max},\,j\in\mathcal{J}},$ the optimizers of \eqref{eq:maxCapBetas}. The constraints guarantee that every feasible set $Z(G^{PE},c\ind{j},\bar{\beta}\ind{j})$ includes a box with semi-edge lengths given by $G\beta\ind{j}_{\max}$. To keep a certain amount  ${\bar{r}=\eta\bar{r}_{\max}}$, ${\eta\in[0,1]}$, of regulation power available, every system reserves a portion ${\eta|\beta\ind{j}_{\max}|}$. The remaining ${\beta\ind{j}_{\ts{rem}}:=\bar{\beta}\ind{j}-\beta\ind{j}_{\max}}$ defines a zonotope within which the baseline $p\ind{j}$ can be chosen freely. The cheapest aggregate baseline is ${p\opt=c\ind{\ts{agg}}+G^{PE}\beta^{(\ts{agg})\ast}}$ with ${\beta^{(\ts{agg})\ast}}$ minimizing
\begin{equation}
\begin{aligned}
& \min\limits_{\beta\ind{\ts{agg}}} & & T\ind{\ts{agg}}(\beta\ind{\ts{agg}})\\
			& \hspace*{2mm} \ts{s.t.} & & -\sum\limits_{j\in\mathcal{J}}{\beta\ind{j}_{\ts{rem}}}\leq\beta\ind{\ts{agg}}\leq\sum\limits_{j\in\mathcal{J}}{\beta\ind{j}_{\ts{rem}}}.
		\end{aligned}
		\label{eq:optiBaselineZ}
\end{equation}
This problem is solved efficiently by the disaggregation algorithm of \sect\ref{s:disaggregation}.

Table \ref{tab:baselineCosts} provides the expected baseline costs $\hat{R}$ for different amounts of regulation power $\bar{r}$ reserved. The costs are computed for a population of 100 PEVs with the parameters sampled from \tab\ref{tab:sysParams}. Historic EPEX day-ahead prices are used for $\hat{v}$. Considered are the cases where i) problem \eqref{eq:defBaseCosts} is solved using the original polytopic feasible sets, problem \eqref{eq:optiBaselineZ} is solved for zonotopic feasible sets, and iii) problem \eqref{eq:optiBaselineZ} is solved for feasible sets approximated by axis-aligned boxes. While the zonotopic approximation is suboptimal compared with the solution found on the original polytopic sets, it still yields significantly lower baseline costs than feasible sets approximated by boxes do. These results indicate that zonotopes are a reasonable compromise between the complex, but accurate polytopic feasible sets and the simple, but inaccurate approximations by boxes.

\begin{table}[tbh]
\setlength{\abovecaptionskip}{-5pt}
\caption{Baseline costs for different types of feasible sets.}
\label{tab:baselineCosts}
\begin{center}
\begin{tabular}{|r|c|c|c|c|c|c|}  \hline
      & ${\bar r=0}$ kW & 20 kW & 40 kW & 60 kW \\ \hline 
	Polytopes & -77.9 EUR & -59.5 EUR & -41.0 EUR & -22.2 EUR\\
	Zonotopes & -77.7 EUR & -58.6 EUR & -39.1 EUR & -18.5 EUR\\
	A-A Boxes & -43.1 EUR & -25.8 EUR & -8.6 EUR  & 8.7 EUR \\\hline
\end{tabular}
\end{center}
\end{table}

%% file: conclusion.tex
\section{Conclusion}
\label{s:conclusion}

The description, aggregation, and disaggregation of the energetic flexibility of energy resources are important and challenging tasks. This paper introduced a generic and scalable approach for flexible energy systems to quantitatively describe and price flexibility based on zonotopic sets. The description can be used to aggregate flexibility efficiently, disaggregate control decisions, and solve aggregate-level problems such as computing the minimum baseline costs under regulation capacity constraints. It has been shown that zonotopic flexibility yields a reasonable compromise between accuracy of approximation and computational complexity.
Future work will include the incorporation of additional features, such as ramp rates, in the zonotope representation. Further, the applicability of distributed optimization methods for solving various disaggregation problems will be investigated.


%% file: appendix.tex
\appendices

\section{Facets of PE-Zonotopes}
\label{a:zonoNormals}
In this section we characterize the facets of zonotopes defined by \eqref{eq:myGen1}-\eqref{eq:myGen2}. It follows that zonotopes can reconstruct every possible facet of a PE-polytope.

\begin{thm}\label{thm:facet1}
Every facet of a full dimensional zonotope ${Z \subseteq \mathbb{R}^N}$ defined by \eqref{eq:myGen1}-\eqref{eq:myGen2} has normal vector ${\sum \{e_i : j\leq i\leq k\}}$ for some ${1 \leq j \leq k \leq N}$
($e_i$ is the unit vector of coordinate $i$).
\end{thm}

\textbf{Proof.}
Let $F$ be a facet of $Z$, $H$ the affine hyperplane generated by $F$, and $c$ a normal vector of $F$ so that ${F=\{z\in Z: zc=M\}}$ with ${M=\max\{zc: z\in Z\}}$. Let ${1 \leq l \leq N}$ be a coordinate which maximizes $|c_l|$. We know that ${c_l \neq 0}$. Let $j$ be the minimum, and $k$ the maximum value in ${\{1,\ldots,N\}}$ for which ${c_i = c_l}$ for all ${j\leq i\leq k}$. $Z$ is centrally symmetric, so without loss of generality we can assume that $c_l>0$. 

The following claim follows from the definition \eqref{eq:myGen1}-\eqref{eq:myGen2} of $Z$.
\begin{clm}\label{facetclm}
Let ${z = c + \sum_{i=1}^g \beta_i g^{(i)} \in Z}$. If ${z\in F}$ then
\begin{enumerate}
\item ${\beta_{N+j-1} = \overline{\beta}_{N+j}}$ if ${j>1}$,
\item ${\beta_{N+k} = -\overline{\beta}_{N+k}}$ if ${k<N}$.
\end{enumerate}
\end{clm}

For a set ${X\subseteq \mathbb{R}^N}$ and ${I\subseteq\{1,\ldots,N\}}$ we denote by $X|_I$ the projection of $X$ onto $\mathbb{R}^I$. Let ${I_1=\{j,\ldots,k\}}$ and ${I_2=\{1,\ldots,N\}-I_1}$. ${I_2=\emptyset}$ is possible. For ${x,y\in F}$ we define ${\gamma(x,y) = x|_{I_1} \times
y|_{I_2}}$, the combination of $x$ and $y$ along the partition ${I=I_1\dot\cup I_2}$. Let ${x=c + \sum \beta^{x}_i g^{i}}$, and ${y=c+ \sum \beta^{y}_i g^{i}}$. By Claim \ref{facetclm} one can write
$\gamma(x,y)$ as ${c + \sum \beta_i g^{(i)} \in F}$ where
\begin{enumerate}
\item ${\beta_i = \beta^{x}_i}$ if ${i\in I_1}$ and ${\beta_i = \beta^{y}_i}$ if ${i\in I_2}$,
\item ${\beta_{N+i} = \beta^{x}_{N+i}}$ if ${i\geq j}$ and ${i<k}$, 
\item ${\beta_{N+i} = \beta^{y}_{N+i}}$ if ${i<j-1}$ or ${i>k}$, 
\item ${\beta_{N+j-1} = \overline{\beta}_{N+j}}$ if ${j>1}$,
\item ${\beta_{N+k} = -\overline{\beta}_{N+k}}$ if ${k<N}$.
\end{enumerate}
It follows that ${\gamma(x,y)\in Z}$. By symmetry, ${\gamma(y,x)\in Z}$. Clearly, ${\gamma(x,y) + \gamma(y,x)=x+y}$ and thus $\gamma(x,y),\,\gamma(y,x)\in F$.

Let ${v_m\in F|_{I_m} \subseteq \mathbb{R}^{I_m}}$ for ${m=1,2}$. By definition, there exists a vector ${z_m = c + \sum \beta_i^m g^{(i)} \in F}$ so that ${z_m|_{I_m}=v_m}$ for ${m=1,2}$. It is easy to see
that ${v_1 \times v_2 = \gamma(z_1,z_2)\in F}$. It follows that ${F =F|_{I_1} \times F|_{I_2}}$, and also that $F|_{I_m}$ is the set of points maximizing $c|_{I_m}$ in $Z|_{I_m}$.

${\dim(F|_{I_1}) \leq |I_1|-1}$ because the vectors $g_{N+i}$ for ${j\leq i\leq k}$ are generators, and clearly ${\dim(F|_{I_2}) \leq |I_2|}$. Now ${N-1 = \dim(F) = \dim(F|_{I_1}) + \dim(F|_{I_2}) \leq |I_1| + |I_2| - 1 = N-1}$, and thus equality most hold throughout. $F|_{I_2}$, being the set of points maximizing $c|_{I_2}$ in $Z_{I_2}$, can only be full-dimensional if ${c|_{I_2}=0}$. After normalizing, we get that $c$ has the form ${\sum \{e_i : j\leq i\leq k\}}$.\qed

\begin{thm}\label{thm:facet2}
Let ${Z \subseteq \mathbb{R}^N}$ be a zonotope defined by \eqref{eq:myGen1}-\eqref{eq:myGen2}. If ${\overline{\beta}>0}$ then for all ${j\leq i\leq k}$ the objective vector ${c=\sum \{e_i : j\leq
i\leq k\}}$ defines a facet.
\end{thm}

\textbf{Proof.}
Let ${F=\{z\in Z: zc=M\}}$ with ${M=\max\{zc: z\in Z\}}$, and $H$ the
affine hyperplane generated by $F$. Clearly, the following vectors are generators of $H$:
\begin{enumerate}
\item $g^{(i)}$ for ${i<j}$ and ${i>k}$, 
\item $g^{(N+i)}$ for ${i\geq j}$ and ${i< k}$.
\end{enumerate}
The rank of this set of vectors is ${N-1}$, so $H$ is a hyperplane and $F$ is a facet.
\medskip

It follows from Theorems \ref{thm:facet1} and \ref{thm:facet2} that
the maximum number of facets of a zonotope is ${N^2+N}$ and this bound
is reached if ${\overline{\beta}>0}$.


